%% file: ms.tex
\documentclass[10pt,twocolumn,letterpaper]{article}

\pdfoutput=1

\usepackage{iccv}
\usepackage{times}
\usepackage{epsfig}
\usepackage{graphicx}
\usepackage{amsmath}
\usepackage{amssymb}

\usepackage[pagebackref=true,breaklinks=true,letterpaper=true,colorlinks,bookmarks=false]{hyperref}

\iccvfinalcopy 

\def\iccvPaperID{6317} 

\ificcvfinal\fi

\begin{document}

\title{Dynamic CT Reconstruction from Limited Views with Implicit Neural Representations and Parametric Motion Fields}

\author{Albert W. Reed$^{1}$\\
{\tt\small awreed@asu.edu}

\and
Hyojin Kim$^{2}$\\
{\tt\small kim63@llnl.gov}

\and
Rushil Anirudh$^{2}$ \\
{\tt\small anirudh1@llnl.gov}

\and
K. Aditya Mohan$^{2}$ \\
{\tt\small mohan3@llnl.gov}

\and
Kyle Champley$^{2}$ \\
{\tt\small champley1@llnl.gov}

\and
Jingu Kang$^{2}$ \\
{\tt\small kang12@llnl.gov} 

\and
Suren Jayasuriya$^{1}$ \\
{\tt\small sjayasur@asu.edu} 


}

\maketitle%

\noindent$^1$School of Electrical, Computer and Energy Engineering; Arizona State University\\
	$^2$Lawrence Livermore National Laboratories \\
\ificcvfinal\thispagestyle{empty}\fi

\begin{abstract}
Reconstructing dynamic, time-varying scenes with computed tomography (4D-CT) is a challenging and ill-posed problem common to industrial and medical settings. Existing 4D-CT reconstructions are designed for sparse sampling schemes that require fast CT scanners to capture multiple, rapid revolutions around the scene in order to generate high quality results. However, if the scene is moving too fast, then the sampling occurs along a limited view and is difficult to reconstruct due to spatiotemporal ambiguities. In this work, we design a reconstruction pipeline using implicit neural representations coupled with a novel parametric motion field warping to perform limited view 4D-CT reconstruction of rapidly deforming scenes. Importantly, we utilize a differentiable analysis-by-synthesis approach to compare with captured x-ray sinogram data in a self-supervised fashion. Thus, our resulting optimization method requires no training data to reconstruct the scene. We demonstrate that our proposed system robustly reconstructs scenes containing deformable and periodic motion and validate against state-of-the-art baselines. Further, we demonstrate an ability to reconstruct continuous spatiotemporal representations of our scenes and upsample them to arbitrary volumes and frame rates post-optimization. This research opens a new avenue for implicit neural representations in computed tomography reconstruction in general.  



\end{abstract}

\input{Sections/Introduction}

\input{Sections/RelatedWork}

\input{Sections/Method}

\input{Sections/Implementation}

\input{Sections/ExperimentalResults}

\input{Sections/Discussion}

\scriptsize{\subsubsection*{Disclaimer}
This document was prepared as an account of work sponsored by an agency of the United States government. Neither the United States government nor Lawrence Livermore National Security, LLC, nor any of their employees makes any warranty, expressed or implied, or assumes any legal liability or responsibility for the accuracy, completeness, or usefulness of any information, apparatus, product, or process disclosed, or represents that its use would not infringe privately owned rights. Reference herein to any specific commercial product, process, or service by trade name, trademark, manufacturer, or otherwise does not necessarily constitute or imply its endorsement, recommendation, or favoring by the United States government or Lawrence Livermore National Security, LLC. The views and opinions of authors expressed herein do not necessarily state or reflect those of the United States government or Lawrence Livermore National Security, LLC, and shall not be used for advertising or product endorsement purposes.
}

\scriptsize{\subsubsection*{Acknowledgment}
This work was performed under the auspices of the U.S. Department of Energy by Lawrence Livermore National Laboratory under Contract DE-AC52-07NA27344.  LLNL-CONF-816780.
}


{\small
\bibliographystyle{ieee_fullname}
\bibliography{egbib}
}

\end{document}

%% file: Sections/Introduction.tex
\section{Introduction}

\begin{figure*}
\centering
\includegraphics[trim={0cm 9cm 10cm 0cm}, clip, width=0.90\textwidth]{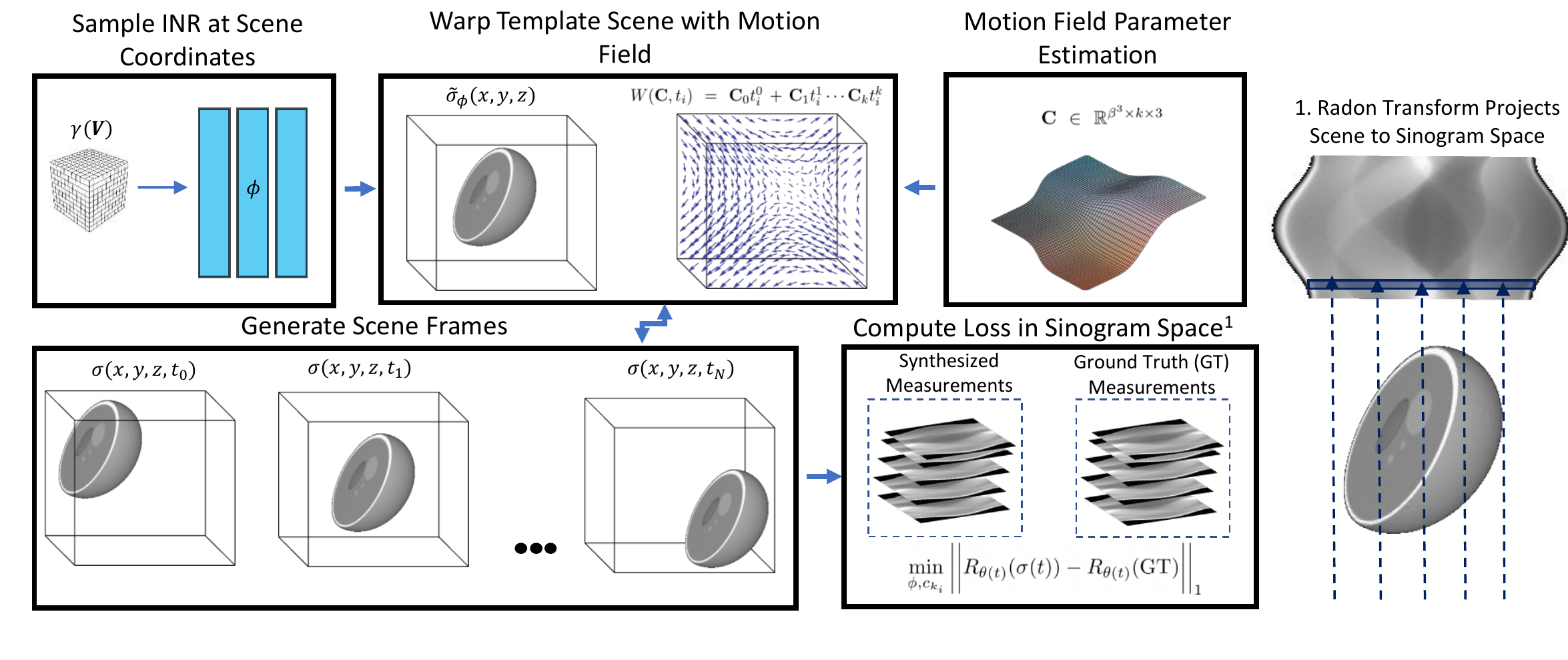}
\caption{Given a sinogram, we jointly estimate a scene template and motion field to reconstruct the 4D scene. Here, we warp a 3D Shepp Logan template to reconstruct its linear translation in time. We simulate sinogram measurements from this 4D scene and compute a loss with the given sinogram. This loss is backpropagated to the implicit neural representation (INR) network weights and motion field parameters until convergence.} 
\label{fig:pipeline}
\end{figure*}

Computed-tomography (CT) is a mature imaging technology with vital medical and industrial applications ~\cite{kachelriess2000ecg, goerres2002pet, de2014industrial}. CT scanners capture x-ray data or \textit{sinograms} by scanning a sequence of angles around an object. Reconstruction algorithms then estimate the scene from these measured sinograms. CT imaging in both 2D and 3D of static objects is a well-studied inverse problem with both theoretical and practical algorithms~\cite{kek, ritschl2011improved, andersen1989algebraic}. 

However, reconstruction of dynamic scenes (\ie scene features changing over time), known as \textit{dynamic 4D-CT}, is a severely ill-posed problem because the sinogram aggregates measurements over time which yields spatio-temporal ambiguities~\cite{gibbs2015three,keall2004acquiring,zang2018space, zang2019warp, mohan2015timbir}. Analogous to motion blur, a static or quasi-static scene is only captured for a small angular range of the sinogram (motion blur analogy: short exposure), and this mapping is a function of the amount of scene motion relative to the CT scanner's rotation speed. Traditional CT reconstruction algorithms have limited capability to address 4D-CT problems due to difficulty accounting for this motion. Yet solving these 4D-CT problems is critically important for a range of applications from clinical diagnosis to non-destructive evaluation for material characteristics and metrology. 

4D-CT reconstruction techniques have been proposed in the literature to handle periodic motion~\cite{pan20044d, keall2004acquiring} as well as more general nonlinear, deformable motion~\cite{zang2018space, zang2019warp, mohan2015timbir}. While the latter methods achieve state-of-the-art for deformable motion, they typically assume slow motion relative to the scanner rotation speed. Specifically, these algorithms assume sparse measurements of the object that span the full angular range ($0-360^\circ$) and typically require multiple revolutions around the sample to reconstruct images at each time step. This approach is called \textit{sparse view} CT in the literature~\cite{niu2014sparse}. However, this setting is not always practical such as when the object motion is too fast for the scanner to make multiple revolutions. Instead, an alternative approach is to collect measurements over partial angular ranges in a \emph{single} revolution -- which is the more challenging \textit{limited-view} reconstruction ~\cite{sidky2006accurate}. Even in the static case, the limited-view problem is challenging due to missing information often leading to significant artifacts \cite{huang2017restoration}. 

\textbf{Key Contributions:} In this paper, we propose a novel, training-data-free approach for 4D-CT reconstruction that works especially well in limited-view scenarios. Our method, illustrated in Figure~\ref{fig:pipeline}, consists of an implicit neural representation (INR) \cite{mildenhall2020nerf} model that acts as the static scene prior coupled with a parametric motion field to estimate an evolving 3D object over time. The reconstruction is then synthesized into sinogram measurements using a differentiable Radon transform to simulate parallel-beam CT scanners. By minimizing the discrepancy between the synthesized and observed sinograms, we are able to optimize both the INR weights and motion parameters in a self-supervised, analysis-by-synthesis fashion to obtain accurate dynamic scene reconstructions without training data.

We leverage recent advances in INRs that use positional encoding to map input coordinates to volume density coefficients. In our experiments, we show that this approach is adept for inverting CT measurements and outperforms conventional methods in reconstruction quality and robustness to noise. Further, since our method learns a continuous representation of both volume and time, we are able to solve the reconstruction problem at low resolutions and then super-resolve our scenes spatio-temporally after the fact. 

\textbf{Validation:} Acquiring 4D-CT data is challenging and one of the primary bottlenecks for research in this area. While this is partly due to the expense and logistics of accessing CT scanners and data, it is also because acquiring ground truth in the 4D case is exceptionally challenging. While there are examples of real CT data ~\cite{zang2018space}, these datasets are specific to certain scanners and specialized, sparse sampling schemes. To address this lack of data and highlight our method's ability to resolve 4D scenes from limited angles, we introduce a synthetic dataset for parallel beam CT. This dataset is generated with an accurate physics simulator for material deformation and used to benchmark ours and competing state-of-the-art methods. We also evaluate our algorithms performance on publicly available thoracic CT reconstructions where we resimulate x-ray measurements. In all cases, we find our method outperforms competitive baselines. 

%% file: Sections/RelatedWork.tex
\section{Related Work}

\textbf{3D-CT Sparse and Limited View Reconstruction:} Traditional CT reconstruction is a mature imaging problem with applications in security, industrial and healthcare. For sparse-view 3D-CT, common techniques include the algebraic reconstruction technique (ART) and the filtered backprojection algorithm (FBP) ~\cite{andersen1984simultaneous, kek}. Model-based approaches have been proposed for the limited angle case ~\cite{huang2017restoration, venkatakrishnan2013model}. More recently, deep-learning based approaches utilize training data ~\cite{han2017framing,kang2016deep,jin2017deep, zhang2016image, venkatakrishnan2013model, anirudh2018lose, kim2019fewview} to estimate scenes from sparse-view or limited angles. For a more comprehensive characterization on limited-angle tomography, we refer the reader to \cite{frikel2013characterization}. In our paper, we are interested in the 4D-CT problem, particularly for limited view sampling.

\textbf{4D-CT for Periodic Motion:} When recovering periodic motion, such as the breathing phases of clinical patients, several methods~\cite{keall2004acquiring, pan20044d, taubmann2017spatio} gate measurements into phase/amplitude cycles to help reconstruct 3D image volumes. This gating limits the number of angular measurements per phase and can induce motion artifacts due to phase error~\cite{schwemmer2013opening}. Similar to our approach, these methods sometimes include parametric motion models for more general non-periodic motion (\eg, heart + breathing motion)~\cite{schwemmer2013opening, rohkohl2008c, schwemmer2013residual}. These motion models perform image registration between estimated phase images, but typically require either enough phase gated measurements or relatively slow motion to work effectively. In contrast, our approach does not require phase information for reconstruction. 




\textbf{4D-CT for Deformable Motion:} State-of-the-art 4D-CT reconstruction methods jointly estimate the scene and motion field parameters ~\cite{zang2018space, mohan2015timbir, zang2019warp} to solve for non-periodic deforming scenes. The authors of ~\cite{mohan2015timbir} show that quickly rotating the CT scanner and sampling sparse angular views in an interlaced fashion enables the capture of high-fidelity images. In ~\cite{zang2018space}, the authors introduce low-discrepency sampling and jointly solve for the motion flow-field and scene volume. This work was extended in~\cite{zang2018space} by using the low-discrepency sampling and adapting the reconstruction algorithm to dynamically upsample temporal frames when the motion is rapidly changing the measured object. Importantly, all these methods require quickly rotating the CT scanner to capture a sparse set of angular measurements that span 360 degrees, a constraint we do not require.

 
 %


\textbf{Implicit Neural Representations (INR): }
Recently, coordinate based multi-layer perceptrons, or INRs, have found success in the imaging domain. These architectures learn functions that map input coordinates (\eg, $(x, y, z)$) to physical properties of the scene (\eg, density at $(x, y, z))$. These networks coupled with differentiable renderers have demonstrated impressive capabilities for estimating 3D scenes ~\cite{mildenhall2020nerf, loper2014opendr, liu2019soft, zhang2020nerf++}. Our method is inspired by~\cite{mildenhall2020nerf}, who estimate a continuous 3D volume density from a limited set of 2D images. More recently, \cite{chen2020learned,chibane2020implicit,dupont2020equivariant,riegler2020free,zhang2021neural} exploit the NeRF architecture to solve problems like view synthesis, texture completion from impartial 3D data, non-line-of-sight imaging recognition, etc. In \cite{park2020deformable}, the authors introduced a method to jointly learn a scene template and warp field for transforming the template through time for RGB video frames. Our work is similar in approach, as we also learn a scene template and then a continuous mapping through time. However our application is wildly different by leveraging 4D-CT measurements and incorporating a differentiable Radon transform into our pipeline. Concurrently at the time of writing this paper, recent work in \cite{sun2021coil} demonstrates the potential of INRs to solve ill-posed inverse problems in the tomographic imaging domain. However, this paper only considers 3D static scenes whereas we consider 4D scenes and under limited view sampling.

%% file: Sections/Method.tex
\section{4D-CT Forward Imaging Model}

The primary task of this paper is to reconstruct 3D scenes in time from CT measurements (\ie sinograms). In this section, we formulate our forward imaging model and the key assumptions our algorithm makes for reconstructing scenes from CT angular projections. Next, in section~\ref{sec:method}, we discuss our algorithmic pipeline to recover 4D scenes (\ie 3D volume in time) from CT sinograms. 

Mathematically, in a parallel-beam CT configuration, the three dimensional Radon transform models the CT measurement of a dynamic scene at a particular viewing angle as follows:
\begin{equation*}
    p_{\theta}(r, z) = \iint \sigma(x, y, z, t) \delta(x\cos(\theta(t)) + y\sin(\theta(t)) - r) dxdy,
\end{equation*}
where $\sigma(x, y, z, t)$ is scene's linear attenuation coefficient (LAC) at coordinates $(x, y, z)$ and time $(t)$, $\theta(t)$ is the view angle at time $t$, and $\delta(\cdot)$ is the Dirac delta function. The result of this transform, $p_{\theta}(r, z)$, is the summation of the scene's LAC at view angle $\theta$, detector pixel $r$ and height $z$. In other words, $\{p_{\theta}(r, z)| (r,z) \in \mathbb{R}^2\}$ is the 2D projection of the 3D volume $\sigma(x, y, z, t)$ at the view angle $\theta$ and time $t$. We assume the scene's LAC remains fixed for single projection, such that $||\sigma(x, y, z, t + dt) - \sigma(x, z, y, t)||_1 = 0$ where $dt$ is the exposure time of the CT scanner. 

\subsection{Modeling Assumptions:}
Here we discuss the main assumptions made by our synthetic dataset and reconstruction algorithm. 

\textbf{Parallel-Beam Geometry:} Our first key assumption is that we assume a parallel-beam geometry, the geometry usually used by synchrotron CT scanners. Synchrotron scanners have vital applications for medical and industrial applications~\cite{pacureanu2012nanoscale, kastner2010comparative,helfen2013laminographic}, and we expect our algorithm to be useful in these domains. Our reconstruction geometry would need to be modified to reconstruct CT measurements from cone-beam or helical measurements, typically used in medical imaging applications. The required modification is replacing our Radon transform with a differentiable volume ray tracer, for example those shown in ~\cite{mildenhall2020nerf, park2020deformable}. We leave this modification of our algorithm to future work. 

\textbf{Data Pre-Processing:} We assume CT measurements have undergone pre-processing to account for beam hardening and truncation. First, we consider the development of our algorithm for synchrotron systems which employ a monochromatic x-ray source and are therefore not affected by beam hardening. In cases where the effect is unavoidable, many methods exist for correcting beam hardened measurements and subsequently enabling high quality reconstructions~\cite{herman1979correction, brooks1976beam}. Truncation is not common in scientific and industrial imaging, but when present in measurements, methods exist for correcting it~\cite{zamyatin2007extension}.

 \begin{figure}[h]
\includegraphics[trim={0cm 4.5cm 7cm 0cm}, clip, width=8cm]{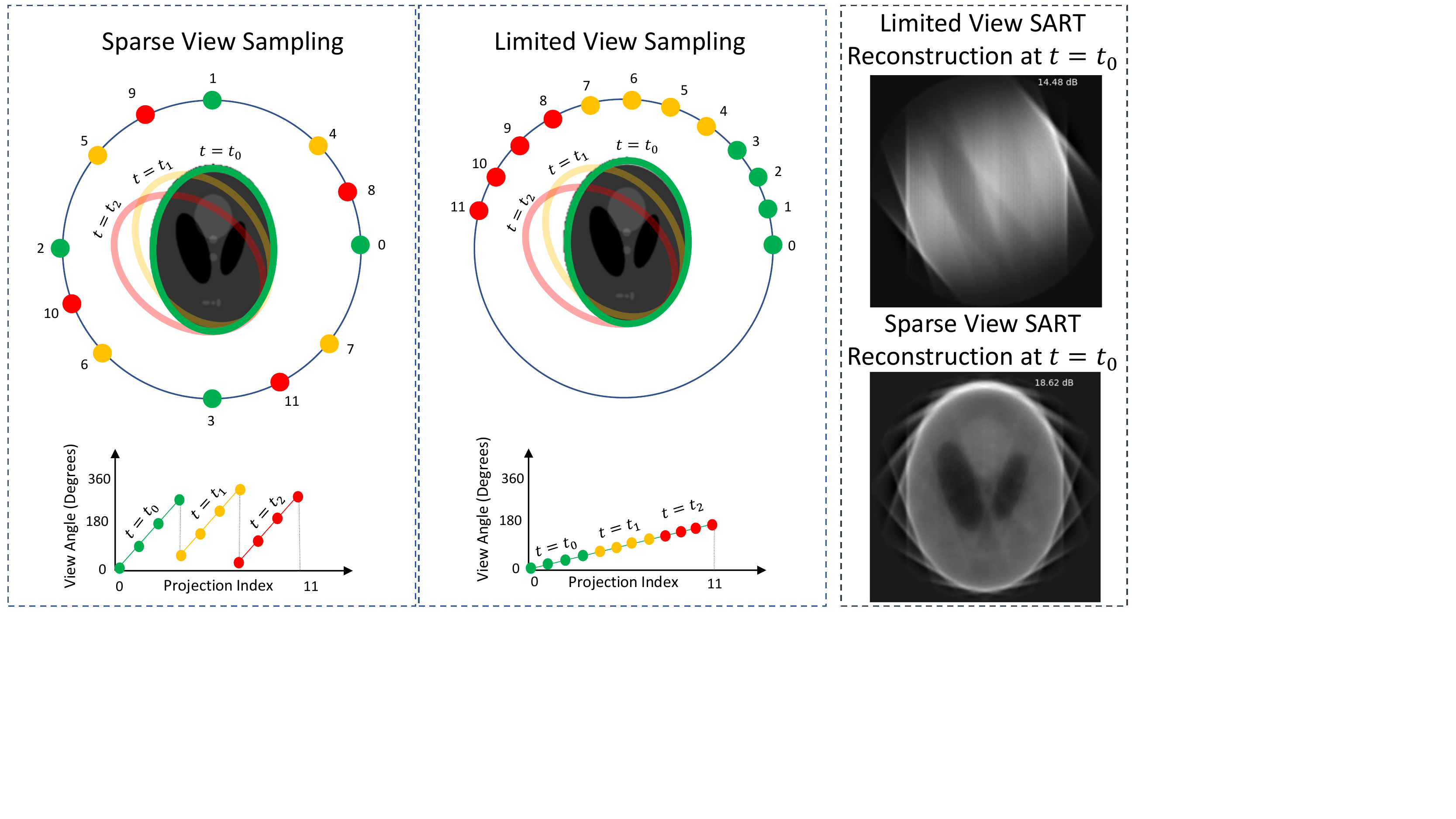}
\caption{Left: Sparse view sampling used by baseline methods ~\cite{mohan2015timbir, zang2018space, zang2019warp}. Right: Limited view sampling scheme we address with our method. Reconstructing objects from limited views is traditionally considered more challenging than from sparse views. We illustrate this fact on the far right of the figure where we show SART ~\cite{andersen1984simultaneous} reconstructions of the Shepp-Logan phantom ~\cite{shepp1974fourier} from $20$ limited views versus $20$ sparse views. Note the superior structural recovery of the phantom under sparse view sampling.}
\label{fig:sampling}
\end{figure}

\textbf{Limited View Sampling:} Previous 4D-CT methods assume either phase-based information or a sparse set of angular samples in order to reconstruct the scene. However, objects undergoing general deformations are not amenable to phase gating and sparse sampling schemes require scanner rotation speed to be fast relative to scene motion. Our method relaxes these assumptions --- we do not require a specialized sampling scheme or phase information for reconstruction. Rather, we assume only that the scene is captured contiguously between start and stop angles and that motion is static in between each projection. Consequently, our sampling method requires fewer revolutions of the turntable during a single scan, and we show that we outperform state-of-the-art methods under this sampling scheme. This fact implies our method will enable slower CT scanners to capture moving objects at a fidelity that was impossible with existing methods. In Figure~\ref{fig:sampling}, we illustrate the differences between limited view and sparse view sampling for CT imaging.

\section{System Architecture}
\label{sec:method}

As described earlier, limited view 4D CT is an ill-posed problem both due to scene motion as well as the incomplete data captured from dense angular sampling. Our key insight is to leverage implicit neural representations to jointly learn continuous functions of the scene volume and its evolution in time. In this section, we present our algorithmic pipeline consisting of three main parts: (1) an INR to estimate a template reconstruction of the static 3D volume LACs; (2) a parametric motion field that warps the template in time; and (3) a differentiable Radon transform to synthesize an estimate of the sinogram measurements. This pipeline is optimized jointly via analysis-by-synthesis with the ground truth sinogram measurements in a self-supervised fashion.

\textbf{Template Estimation:} In order to estimate a template of our measured volume's LACs, we utilize a implicit neural representation architecture, as illustrated in the upper left portion of Figure~\ref{fig:pipeline}. We denote the tunable parameters of the MLP as $\phi$. Specifically, we use a multi-layer perceptron (MLP), denoted by the function $\tilde{\sigma}_\theta$, that maps scene coordinates $(x, y, z)$ to a template reconstruction of the scene's LACs $\tilde{\sigma}(x, y, z)$, such that $\tilde{\sigma}_\theta: (x, y, z) \mapsto \tilde{\sigma}(x, y, z)$. Note that $\tilde{\sigma} \neq \sigma$, \ie, this template reconstruction does not equal the actual reconstruction of the scene's LACs until we use the parametric motion field to warp the template to be consistent with the measurements given in the sinogram. In implementation, we input a grid of $(x, y, z)$ coordinates. Specifically, let $\mathbf{V} \in \mathbb{R}^{\beta^3 \times 3}$ be a voxel representation of the scene, and the scene boundaries defined as $[-1, 1]^3$. For our experiments, we set $\beta = 80$ and sample our INR with $80^3$ linearly spaced coordinates at each iteration. Importantly, we perturb these coordinates randomly within their voxel so the INR is forced to learn a continuous representation of the scene. 

Our INR architecture is inspired by~\cite{mildenhall2020nerf} --- we use 4 fully MLP layers with ReLU activiations. We utilize Gaussian random Fourier feature (GRFF)~\cite{tancik2020fourier} to randomly encode input coordinates with sinusoids of a random frequency. Formally, let $\mathbf{v} = (x,y,z)$ be a coordinate from the input grid. Its GRFF is computed as $\gamma(\mathbf{v}) = [\cos{(2\pi \kappa \mathbf{B} \mathbf{v})}, \sin{(2\pi \kappa \mathbf{B} \mathbf{v})}]$, where $\cos$ and $\sin$ are performed element-wise; $\mathbf{B}$ is a vector randomly sampled from a Gaussian distribution $\mathcal{N}(0,I)$, and $\kappa$ is the bandwidth factor which controls the sharpness of the output from the INR. Similar to~\cite{tancik2020fourier}, we find that tuning the $\kappa$ parameter regularizes our reconstruction. As shown in the supplemental material, setting $\kappa$ too low prevents the INR from fitting high frequency content in the scene. Conversely, setting it too high causes the INR to fit spurious features in the measured sinogram resulting in poor reconstruction quality.

 \textbf{Motion Estimation:} To map the estimated template LAC to the sinogram measurements, we introduce a parametric motion field to warp the template to different time values (\ie, $\tilde{\sigma} \to \sigma(x,y,z,t_0), \sigma(x,y,z,t_1), \dots, \sigma(x,y,z,t_N)$). Specifically, we define a tensor $\mathbf{C} \in \mathbb{R}^{\beta^3 \times k \times 3}$. This tensor contains $k$ polynomial coefficients at each scene voxel in $\beta^3$ in the $3$ spatial dimensions $(x, y, z)$. Next, we define $N$ time samples linearly spaced within $[0, 1]$ where $N$ is the number of angular measurements as $t_{i=0 \cdots N-1}$. To warp a voxel to a specific time $t_i$, we compute the polynomial $W(\mathbf{C}, t_i) = \mathbf{C}_0t_i^0 + \mathbf{C}_1t_i^1 \cdots \mathbf{C}_kt_i^k$, where $W(\mathbf{C}, t_i) \in \mathbb{R}^{\beta^3 \times 3}$ is the warp field of the scene at $t_i$, and $\mathbf{C}_k$ is $\mathbf{C}(:, k, :)$. We generate scene frames using a differentiable grid sampling function as introduced in ~\cite{paszke2019pytorch}, \textit{warp\_fn}$\left(W(\textbf{C}, t_i), \hat{\sigma}\right) = \sigma(x, y, z, t_i)$. Typically, we observe that polynomials of order $k=5$ are sufficient for describing the deformable and periodic motion present in our empirical studies.
 
\textit{Hierarchical motion model:} We observed that attempting to estimate the warp field at the full volume of $\beta^3$ results in poor motion reconstruction. To address this issue, we introduce a hierarchical coarse-to-fine procedure for estimating motion. Specifically, our initial motion field at a base resolution $\alpha$ such that $\mathbf{C}_{\alpha} \in \mathbb{R}^{\alpha^3 \times k \times 3}$ where $\alpha < \beta$. We iteratively increase $\alpha$ throughout training (\eg $2^3 \to 4^3 \to 8^3 \to 16^3 \to \dots$) and use linear upsampling to progressively grow our warp field like $\mathbf{C}_{\alpha_{i + 1}} = U(\mathbf{C}_{\alpha_{i}})$ where $U : \mathbb{R}^{\alpha_i^3} \mapsto \mathbb{R}^{\alpha_{i+1}^3}$. This strategy encourages our optimization to first recover simple motion and then iteratively recover more complex deformations.  

\textbf{Differentiable Radon Transform:}
After estimating a sequence of LAC volumes $\sigma(t_0),\dots, \sigma(t_N)$ by applying our motion field $W$ to the LAC volume template $\tilde{\sigma}$, we then project the LAC volumes through the CT forward imaging model to synthesize CT measurement using the 3D Radon transform. Thus we can compare our synthesized measurements to the ground truth CT measurements provided by the captured sinograms. To enforce a loss, we implement the 3D Radon transform in an autograd enabled package so that the intensity of each projected pixel is differentiable with respect to the viewing angle. We backpropagate derivatives through this operation and update the INR and motion field parameters for analysis-by-synthesis.


The weights of the INR and the coefficients of the parametric motion field are updated via gradient descent to minimize our loss function 
\begin{equation}\label{eq:4}
\begin{aligned}
    \min_{\phi, \mathbf{C}} \lambda_{1}\big |\big| R_{\theta(t)}(\sigma(x, y, z, t) - R_{\theta(t)}(GT)\big|\big|_1 \\
    +  \lambda_{2} \mathrm{TV}(\mathbf{C}), \quad t \in [0, 1]. 
\end{aligned}
\end{equation}
The first term in our loss function is an L1 loss between our synthesized and given sinogram measurements. Additionally, we regularize the motion field by penalizing the spatial variation of the coefficients. The weight of $\lambda_2$ governs the allowed spatial complexity of the motion field. Higher values create smoother warp fields but may underfit complex motion, and low values allow fitting of complex motion but are more prone to noisy solutions.

\textbf{Continuous Volume and Time Representation:}
\label{cont_sec}
Due to memory constraints, we sample our INR at resolution of $80^3$ points during optimization. However, similar to ~\cite{mildenhall2020nerf}, we randomly perturb these points at each iteration, encouraging our INR to learn a continuous mapping from $(x, y, z)$ to scene LAC. This continuous mapping property is useful because it allows us to query our INR at an arbitrary resolution post-optimization. Similarly, we sample our motion field at random times $t$ during the optimization which encourages the polynomial coefficients to fit a continuous representation of the scene motion. We use this fact to upsample our scenes to arbitrary frame rates post-optimization. Due to the parameteric representation of the motion field, we are constrained to simple trilinear interpolation for upsampling the field itself. We find that this works well in practice, but is something to be addressed in future work.

Using the upsampling functionality, we show that we can optimize our scene on a set of relatively low-resolution measurements (\eg $80^3$) at $10$ time frames, and then upsample the measurements to $256^3$ at $90$ time frames ($256\times256\times256\times90$), making our method practically viable while also bypassing substantial GPU memory requirements. We demonstrate results of this fact in Section ~\ref{results_sec} and videos in the supplement.

%% file: Sections/Implementation.tex
\section{Implementation}

\begin{figure*}[!htb]
    \centering
    \includegraphics[width=0.9\textwidth]{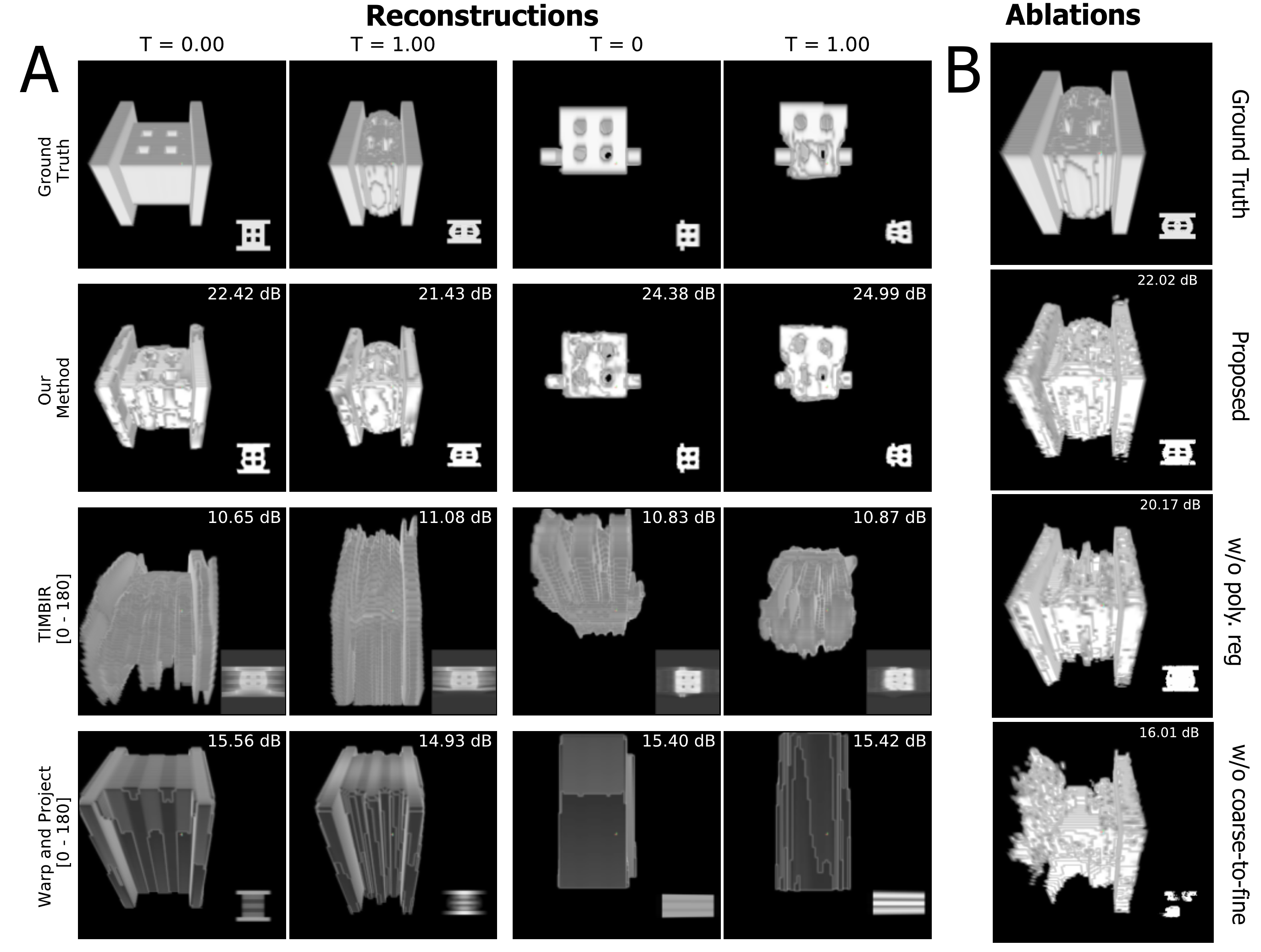}
    \caption{\textbf{(A)} Reconstruction results of our method and competing baseline methods on two objects (Left: Object Alum\#1. Right: Object Alum\#2.) from our aluminum deformation dataset at the beginning ($T=0.00$) and end ($T=1.00$) of the deformation. On the left, two plates compress the center mass of the object, and on the right, the center of the object is squeezed by two bars. In each tile, we display 3D rendering of the object to the left, and on the right, a white 2D inlet containing an $XY$ slice through the center of the object. The PSNR of each frame is shown in white at the upper right corner of each method's tile. \textbf{(B)} Ablation of key motion field regularization components.}
    \label{fig:deformable_results}
\end{figure*}

\textbf{Datasets: }We benchmark our algorithm and competing SOA methods on a dynamic 4D-CT dataset of object deformation that we created (D4DCT Dataset). This dataset represents a time-varying object deformation to demonstrate damage evolution due to mechanical stresses over time for the study of materials science and additive manufacturing. The deformation by the damage evolution provides crucial information about the performance and safety of the material of interest, more accurate, physics-based simulation is needed. To this end, we generated dataset using the material point method (MPM)~\cite{MPM_CPDI2} to accurately represent deformation of a type of aluminum under various loading conditions. We then simulated 4D sinogram data with the provided angular ranges of $180$ and $720$ degrees where the number of uniformly spaced projections is $90$ and the detector row size is $80$. The dimension of the ground truth volume is $80^{3}$ and the number of ground truth frames for algorithm evaluation is $10$. We plan to open source this dataset to encourage reproducibility in 4D-CT research. 

We also benchmark our algorithm on thoracic CT data ~\cite{castillo2009framework}. This dataset contains volumetric reconstructions of the chest cavity at $10$ breathing phases. There is motion present due to periodic functions of the diaphragm and heart. We project the 10 reconstructions from an $80^3$ volume into sinogram space with $90$ uniform angular projection between $0$ and $180$ degrees to emulate the real sinogram data. 




\textbf{Comparisons: } To our knowledge, we are the first method to propose solving 4D CT in the limited angle regime without the use of motion phase information. However, we benchmark against two baseline methods typically used for sparse angular views on our limited angle datasets: TIMBIR~\cite{mohan2015timbir} and Warp and Project ~\cite{zang2018space}. TIMBIR uses sparse angular views with interleaved sampling to recover 4D-CT reconstructions. Warp and Project jointly solve for motion and the object reconstruction from sparse angular views. We note that these methods are expected to perform poorly in our limited angle sampling regime as both are designed for sparse sampling in time. We also benchmark the datasets with the filtered back projection (FBP) method for static 3D CT. This method is not designed to account for motion and serves to illustrates the degrading effects motion has on reconstruction performance. We note that we utilize source code for both~\cite{mohan2015timbir} and ~\cite{zang2018space} given to us by the authors for running our experiments. 





\textbf{Algorithm Implementation Details: } Our algorithm is implemented in PyTorch to run on two Titan X GPUs for 15 minutes per recovery of an $80^3$ LAC volume in time from a given sinogram. We use a Fourier Feature value of $\kappa = 1$ or $\kappa = 1.5$, as well as the ADAM optimizer~\cite{kingma2014adam} with learning rate .001, $\lambda_{1} = 1$, and $\lambda_{2} = 0.001$ for all experiments. In the supplemental material, we present a detailed list of network layers and parameters for full reproduction. We run TIMBIR on a one P100 GPU for an average of 2 minutes per $80^3$ scene. Warp and Project ran on a laptop with 16 GB of RAM and ran for approximately 6 hours to reconstruct $80^3$ scenes in our dataset.

%% file: Sections/ExperimentalResults.tex
\section{Experimental Results}
\label{results_sec}

\begin{table}
\centering
\small
\begin{tabular}{| l | l | l | l | l |}
\hline
Object& Ours&\footnotesize{TIMBIR}\cite{mohan2015timbir}&Warp\cite{zang2019warp}&FBP\\
\hline
\small{Alum\#1} & \textbf{22.68/0.95} & 10.95/0.08 & 15.12/0.72  & 9.29/0.08   \\
\small{Alum\#2}& \textbf{24.50/0.96} & 10.74/0.04   & 14.22/0.65 & 10.65/0.07   \\
\small{Alum\#3}& \textbf{26.47/0.98} & 11.23/0.06   & 16.01/0.76 & 14.82/0.12   \\
\small{Alum\#4}& \textbf{26.01/0.98} & 11.06/0.04   & 15.65/0.77 & 9.76/0.06   \\
\small{Alum\#5}& \textbf{26.56/0.98} & 11.39/0.06   & 16.31/0.76 & 13.00/0.11   \\
\small{Alum\#6}& \textbf{24.66/0.97} & 10.98/0.04   & 15.37/0.72 & 9.03/0.05   \\
Thorax & \textbf{22.45/0.90} & 14.82/0.61 & 8.27/0.17   & 15.36/0.63  \\
\hline
\end{tabular}
\caption{Summary of the results shown in Figures ~\ref{fig:deformable_results} and ~\ref{fig:heart_results} benchmarking our algorithm against two SOA methods and the FBP method. We report PSNR/SSIM metrics averaged over $10$ reconstructed estimated and ground truth frames. Alum \#3-\#6 results are shown in the supplemental material.}
\label{table:results}
\vspace{-10pt}
\end{table}

\begin{figure}[!t]
    \centering
    \includegraphics[trim={0cm 9cm 20cm 0cm}, clip, width=\columnwidth]{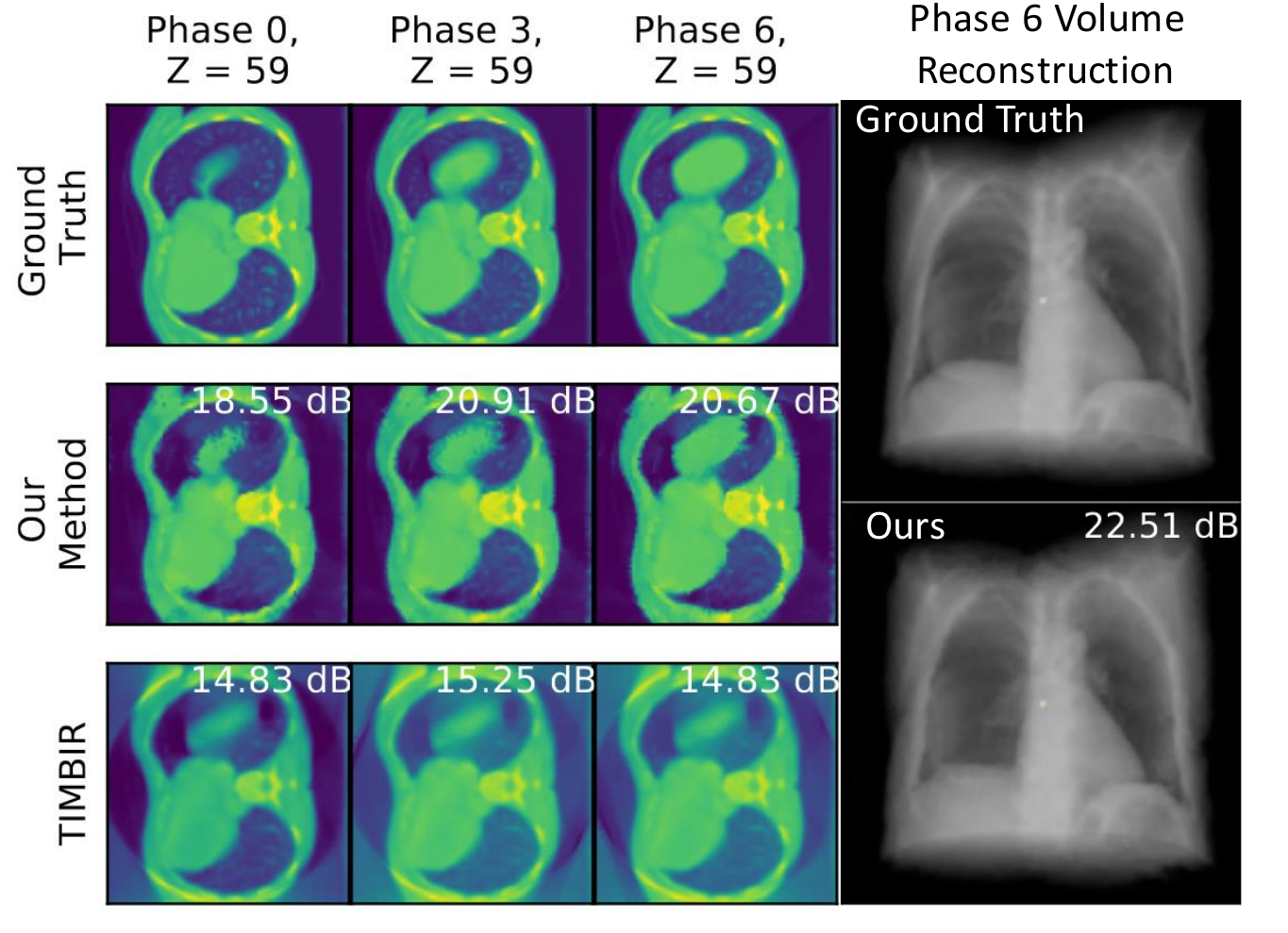}
    \caption{The top row shows ground truth $XY$ slices of the thoracic cavity at three different breathing phases. We show results of our method and TIMBIR in the middle and bottom rows respectively. On the right, we show our volumetric reconstruction of the chest cavity. Please see supplemental materials for a video of this reconstruction.}
    \label{fig:heart_results}
    \vspace{-15pt}
\end{figure}

\begin{figure*}[t]
    \centering
    \includegraphics[width=0.90\textwidth]{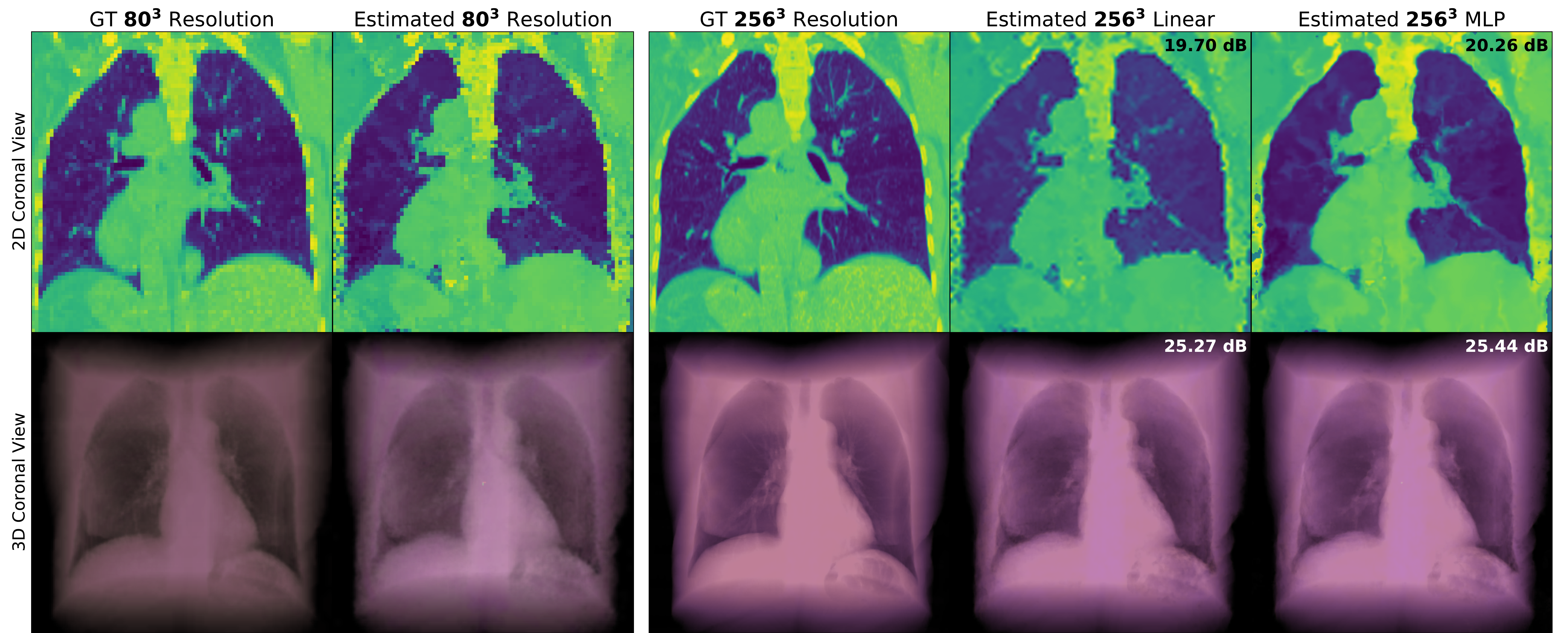}
    \caption{We demonstrate an ability to effectively upsample our scenes to arbitrary resolutions by sampling our INR's continuous representation of the scene. In this experiment, we use the thoracic data to $80^3$ as ground truth (first column), and performed scene reconstruction with our method (second column). In the third column, we show the ground truth thoracic data at its native resolution $256^3$. In the last two columns from left to right, we show the results of upsampling our $80^3$ reconstruction with a trilinear interpolation, and upsampling by querying our INR at an upsampled set of input $(x, y, z)$ coordinates.}
    \label{fig:upsampled}
\end{figure*}



Here we benchmark the performance of our algorithm against baselines on our D4DCT dataset and the thoracic dataset. While we show these results at a resolution of $80^3$ to compare with our baselines, we also demonstrate our ability to upsample our results to a more practical CT resolution of $256^3$. In addition, we show the ability to upsample our videos to arbitrary frame rates, but refer the reader to the supplemental videos for an example of this capability. Finally, we show ablations on our method and and its sampling schemes at the end of this section. 

\textbf{D4DCT Dataset:} As shown in Figure~\ref{fig:deformable_results}(A) and summarized in Table~\ref{table:results}, our proposed method drastically outperforms competing methods in peak signal-to-noise ratio (PSNR) and structural similarity index (SSIM) on the D4DCT dataset. Our method recovers the geometry and deformation of the aluminum object accurately in time. We encourage the reader to refer to the supplemental videos of these reconstructions to see the full deformations over time. 

We observe that our method is not without flaws --- our reconstructions contain artifacts on the surface of the aluminum object. However, we notice that the competing methods of Figure~\ref{fig:deformable_results}(A) contain much more severe artifacts. These artifacts exist because the limited view sampling scheme prevents these methods from constructing sufficient initial estimates of the object at each time step; these algorithms expect a sparse set of samples in order to form these initial reconstructions, as shown in ~\ref{fig:sampling}. We believe our method performs significantly better in this sampling scheme because we optimize a single reconstruction that is warped through time, meaning our optimization leverages the full angular range to reduce artifacts.

\textbf{Thoracic CT Data~\cite{castillo2009framework}:}
In Figure~\ref{fig:heart_results}, we display our reconstruction results for $3$ breathing phases of our thoracic CT data. In the upper half of each transverse slice, the top portion of the diaphragm is observed progressively rising and occupying more of the scene at each breathing phase. Our method recovers this motion and the overall geometry of the thoracic cavity. We encourage the reader to view the supplemental material videos to view the full reconstructions in time. Compared to the TIMBIR reconstruction, we observe that our method preserves sharper details and achieves a better estimate of the motion. We also benchmarked the Warp and Project method on this data, but the reconstruction quality was subpar as noted quantitatively in Table~\ref{table:results}. This may be due to the fact that its code implementation is optimized for sparse angular sampling and not robust to limited angular sampling. 

\begin{figure}[!htb]
    \centering
    \includegraphics[width=0.45\textwidth]{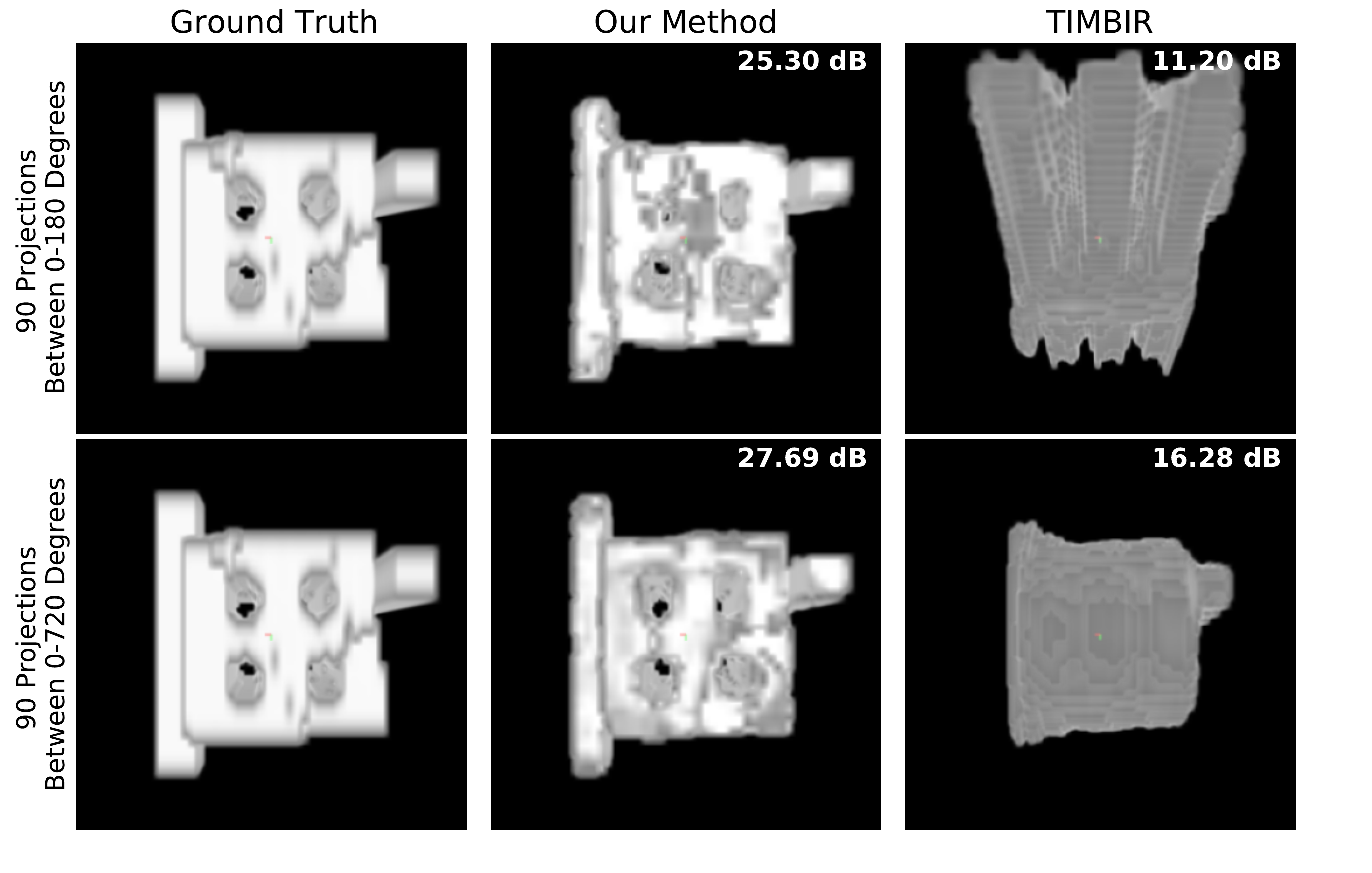}
    \caption{More sparsely sampled projections results in better reconstruction performance for ours and SOA method TIMBIR~\cite{mohan2015timbir}. Here we show the PSNR and visual difference of reconstructing with $90$ uniform projections between $180$ degrees (top row) and between $720$ degrees (bottom row).}
    \label{fig:180vs720}
\end{figure}

\subsection{Ablation Studies}

\begin{figure}[h]
    \centering
    \includegraphics[trim={0cm 4.5cm 15cm 0cm}, clip, width=0.8\columnwidth]{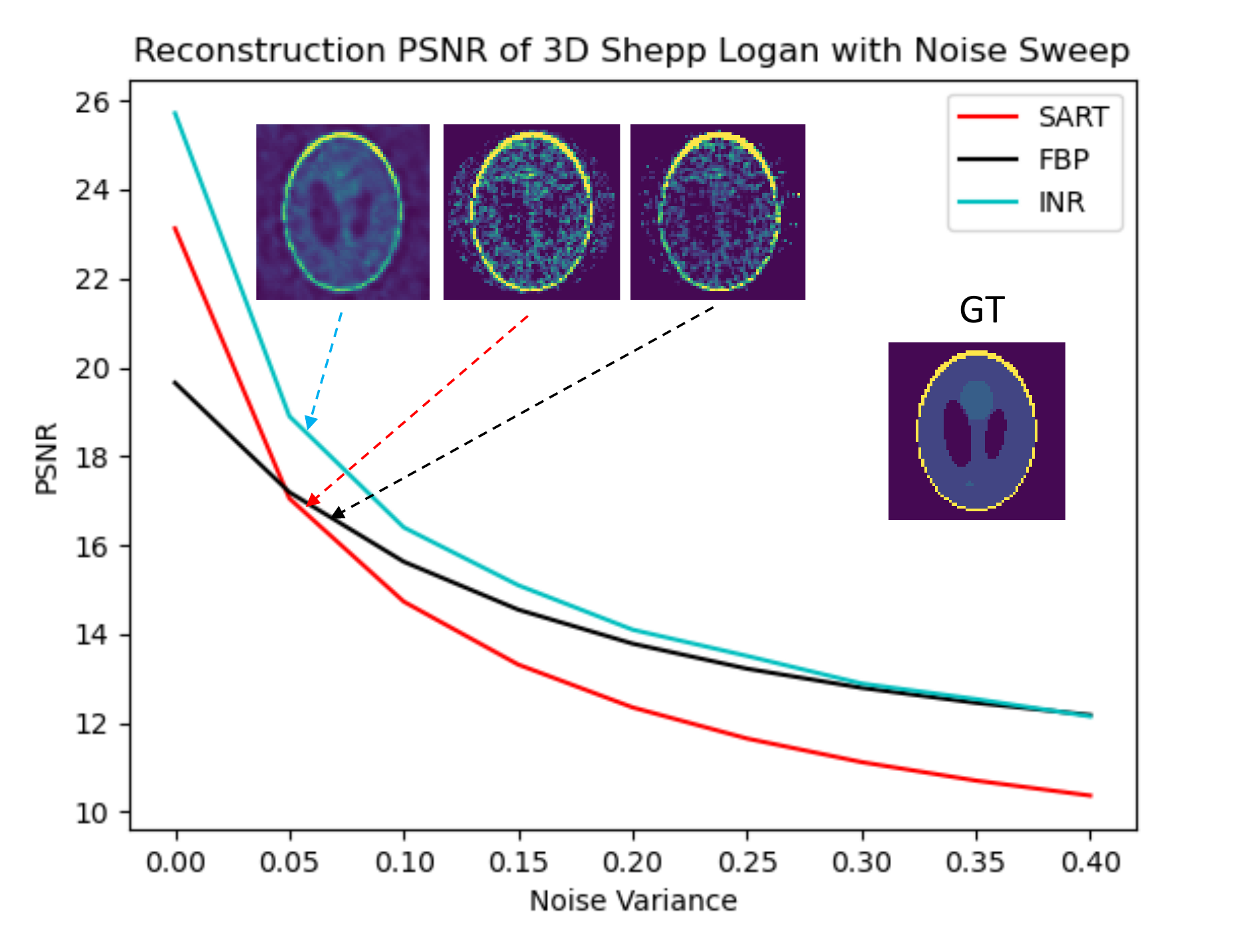}
    \caption{The INR model outperforms conventional CT reconstruction methods ~\cite{kek, andersen1984simultaneous} in static 3D reconstructions. We show transverse slices of the 3D Shepp-Logan for viewer reference.}
    \label{reconstruction_results}
    \vspace{-10pt}
\end{figure}

\textbf{INR Reconstruction:} Using an INR in our reconstruction pipeline allows us several key advantages. First, we observe that our INR outperforms conventional reconstruction methods on 3D scene reconstructions. In Figure ~\ref{reconstruction_results}, we tasked an INR and the two conventional reconstruction methods (SART and FBP), the 3D reconstruction technique used by our 4D-CT baselines~\cite{zang2018space, mohan2015timbir}, to reconstruct a 3D Shepp-Logan phantom. The INR is implemented as a MLP and takes GRFF features of $(x, y)$ coordinates as input to predict the LAC at each coordinate. The scene is projected to sinogram space with the Radon transform and compared to the given limited view measurements to enforce a loss via gradient updates of the INR weights. We observed that the INR gave better reconstruction PSNR. We also tested this performance in the presence of additive noise and still observe better performance under these conditions. We believe this performance gap extends to the 4D problem since our 4D reconstruction results drastically outperform baselines methods that use SART and FBP. 

Secondly, our INR allows us to upsample the scene to arbitrary resolutions in the post-optimization, as shown in Figure ~\ref{fig:upsampled}. Impressively, our INR yields sharp results that resemble ground truth at the high resolution of $256^3$ despite being optimized on data at the low resolution of $80^3$. Further, we observe that it qualitatively and quantitatively outperforms naive trilinear upsampling methods. However, our method is not perfect and fails to capture fine details like arteries at the high resolution. This is possibly because these subtle details were too degraded at the optimization resolution $80^3$ for the INR to recover this structure. 

\textbf{Parametric Motion Field Regularization:}
We regularize our motion field both by choosing an appropriate order for its polynomial equation and with the hierarchical motion model. In Figure~\ref{fig:deformable_results}(B), we illustrate the importance of these methods on the final reconstruction quality of our scenes. Of these two methods, the coarse-to-fine ablation affected the largest change on the reconstruction PSNR with a $6$ dB performance improvement. This result is expected as the motion field recovery is extremely ill-posed --- simplifying its initial estimation ensures the motion field does not immediately overfit to noisy solutions. For the warp field polynomial, we observed that a parametric motion field with low order polynomials underfits non-linear motion, and thus required order 3 or higher for satisfactory performance.

\textbf{Effects of Angular Sampling on Performance: }
We observe enhanced reconstruction performance of our tested methods when we increase the angular range of our projections (\ie, make the samples more sparsely situated). In Figure~\ref{fig:180vs720} we show the reconstruction results from our method and TIMBIR~\cite{mohan2015timbir} on data captured with $90$ projections within $180$ and $720$ degrees. Our method resolves the object geometry and deformation in both cases, whereas TIMBIR only begins to capture the underlying geometry in the latter case.

%% file: Sections/Discussion.tex
\section{Discussion}
We demonstrate that our proposed algorithm outperforms SOA methods in reconstructing limited view 4D-CT measurements of deformable motion. These results have the potential to enable CT scanners to measure rapidly moving scenes with a fidelity that was previously unattainable. Generally, this research has the potential to enable more efficient CT scans in industrial and clinical settings. We plan to open source both our code and the D4DCT dataset for reproducible research (after paper acceptance).

We also address two limitations of our work. First, we only consider a parallel-beam scanning geometry. While this makes our method directly applicable to synchrotron scanners, our method needs to be modified to reconstruct cone-beam data. Several other works provide implementations of differentiable ray tracers capable of modeling this geometry ~\cite{syben2019pyro, gao2020generalizing}, but we leave this modification to future work. Second, we show promising results of upsampling our scenes post-optimization. However, the efficacy of this upsampling needs to be further explored and compared with running the optimization at the full resolution. Even so, we believe the upsampling results we show are a promising method for achieving super-resolution in memory-hungry regimes --- a few very recent works also show impressive supper-resolution results with INRs ~\cite{skorokhodov2020adversarial, sun2021coil}. We hope our work sparks interest in dynamic 4D-CT reconstructions that leverage INRs in the future. 
